\documentclass[twocolumn,twocolumn]{IEEEtran}
\usepackage[T1]{fontenc}
\usepackage{color}
\usepackage{array}
\usepackage{float}
\usepackage{mathrsfs}
\usepackage{amsmath}
\usepackage{amssymb}
\usepackage{graphicx}

\usepackage[unicode=true,
 bookmarks=true,bookmarksnumbered=true,bookmarksopen=true,bookmarksopenlevel=1,
 breaklinks=false,pdfborder={0 0 0},pdfborderstyle={},backref=false,colorlinks=false]
 {hyperref}
\hypersetup{pdftitle={Your Title},
 pdfauthor={Your Name},
 pdfpagelayout=OneColumn, pdfnewwindow=true, pdfstartview=XYZ, plainpages=false}

\makeatletter

\floatstyle{ruled}
\newfloat{algorithm}{tbp}{loa}
\providecommand{\algorithmname}{Algorithm}
\floatname{algorithm}{\protect\algorithmname}

\usepackage[caption=false,font=footnotesize]{subfig}
\usepackage{algorithm}
\usepackage{algorithmic}

\usepackage{multirow} 
\usepackage{amsmath} 
\usepackage{xcolor}

\allowdisplaybreaks[4]

\ifCLASSOPTIONcompsoc
\usepackage[caption=false,font=normalsize,labelfont=sf,textfont=sf]{subfig}
\else
\usepackage[caption=false,font=footnotesize]{subfig}
\fi

\usepackage{cite}
\usepackage{bm}
\usepackage{algorithmic}
\usepackage{algorithm}
\usepackage{graphicx}
\IEEEoverridecommandlockouts

\usepackage{lettrine}

\usepackage{geometry}
\@ifundefined{showcaptionsetup}{}{%
 \PassOptionsToPackage{caption=false}{subfig}}
\usepackage{subfig}
\makeatother

\begin{document}

\title{\textcolor{black}{SFC Deployment in Space-Air-Ground Integrated Networks Based on Matching Game}}

\author{\IEEEauthorblockN{Yilu Cao$^{\dagger}$, Ziye Jia$^{\dagger\ast} $, Chao Dong$^{\dagger}$, Yanting Wang$^{\ddagger }$, Jiahao You$^{\dagger}$, and Qihui Wu$^{\dagger}$\\}
\IEEEauthorblockA{$^{\dagger}$The Key Laboratory of Dynamic Cognitive System of Electromagnetic Spectrum Space, Ministry of Industry and Information Technology, Nanjing University of Aeronautics and Astronautics, Nanjing, Jiangsu, 210016, China\\
$^{\ddagger }$School of software, Northwestern Polytechnical University, Xi'an, Shaanxi, 710129, China\\
$^{\ast}$Email: jiaziye@nuaa.edu.cn}
}

\maketitle
\pagestyle{empty} 

\thispagestyle{empty}
\begin{abstract}
    The space-air-ground integrated network (SAGIN) is dynamic and flexible, which can support transmitting data in environments lacking ground communication facilities. However, the nodes of SAGIN are heterogeneous and it is intractable to share the resources to provide multiple services. Therefore, in this paper, we consider using network function virtualization technology to handle the problem of agile resource allocation. In particular, the service function chains (SFCs) are constructed to deploy multiple virtual network functions of different tasks. To depict the dynamic model of SAGIN, we propose the reconfigurable time extension graph. Then, an optimization problem is formulated to maximize the number of completed tasks, i.e., the successful deployed SFC. It is a mixed integer linear programming problem, which is hard to solve in limited time complexity. Hence, we transform it as a many-to-one two-sided matching game problem. Then, we design a Gale-Shapley based algorithm. Finally, via abundant simulations, it is verified that the designed algorithm can effectively deploy SFCs with efficient resource utilization.
\end{abstract}

\newcommand{\CLASSINPUTtoptextmargin}{0.8in}

\newcommand{\CLASSINPUTbottomtextmargin}{1in}

\section{Introduction}

\lettrine[lines=2]{A}{s} a promising technology in the sixth generation (6G) communication system, the space-air-ground integrated network (SAGIN) is multi-layer and highly heterogeneous\cite{8368236}. SAGIN can be applied to multiple tasks in the scenario such as natural disasters, which lack available ground communication facilities. However, the resources in SAGIN are heterogeneous and cannot be directly shared by various tasks, so the resource utilization is inefficient. Therefore, the flexibility and reconfiguration of SAGIN are required to improve the performance. 

The network function virtualization (NFV) is a technology proposed in recent years\cite{7534741}. Traditional network functions are deployed in the form of specialized hardware devices, while NFV can transforms them into virtual network functions (VNFs) running on general equipments. Hence, we consider introducing NFV into SAGIN to efficiently leverage various resources and capabilities of different nodes. NFV allows VNFs to dynamically migrate from one node to another, which increases the flexible deployment of network functions for various tasks. A task may have multiple VNFs, forming a service function chain (SFC) in sequence. Each SFC can be perceived as a virtual path connected by one or more VNF instances in a certain order. The deployment of SFC provides a new approach for SAGIN resource management.

The deployment of SFC is abundantly studied in the terrestrial networks. However, the researches on using SFCs in the SAGIN system are still in its early stage. For instance, in\cite{9062531}, the authors propose the SFC deployment model for the SAGIN system. Then, they use the high-altitude platform as an example to implement the deployment of SFCs by a heuristic greedy algorithm. Zhou \emph{et al.} \cite{8700139} study SFC for agile task offloading in SAGIN. In\cite{9505612}, the heterogeneous resource orchestration of SAGIN is modeled as a multi-domain virtual network embedding problem, solved by a deep reinforcement learning method. Jia \emph{et al.} \cite{9406391} investigate the VNF orchestration problem in low earth orbit (LEO) satellite networks. Then, they propose a  branch-and-price algorithm to address the optimization problem. In\cite{9375493}, the authors propose a SAGIN architecture for Internet of vehicles based on software-defined networking and NFV. However, as far as the authors' knowledge, the research on SFC deployment in SAGIN is not receiving sufficient attentions. Furthermore, the time-varying characteristics of resources are not studied sufficiently.

Therefore, we construct the system model of SAGIN and leverage the reconfigurable time expansion graph (RTEG), which can express the state of different nodes and connections among these nodes. The same node in different time slots can be regarded as distinct nodes. Based on RTEG, we formulate an optimization problem, aiming to maximize the number of task requirements for SFC deployment while satisfying deployment constraints, resource constraints, and flow constraints. It is a mixed integer linear programming (MILP) problem and hard to solve in limited time complexity. Hence, we propose a multi-slot matching game based algorithm, which incorporates the Dijkstra's algorithm to obtain the shortest path. Finally, simulation results show that the proposed algorithm can effectively obtain better results than the compared methods.

The rest of this paper is arranged as follows. The system model is introduced in Section \ref{sec:System-Model}. In Section \ref{sec:Problem-Formulation}, we formulate the task maximization problem for completing SFC deployment. Section \ref{sec:Algorithm-Design} proposes an effective matching algorithm to deal with the formulated problem. Simulations are provided in Section \ref{sec:Simulation-Results}, followed by the conclusions in Section \ref{sec:Conclusions}.

\section{System Model\label{sec:System-Model}}

\subsection{Network Model}

\begin{figure}[!t]
    \centering
    \centerline{\includegraphics[height=7.85cm,width=8cm ]{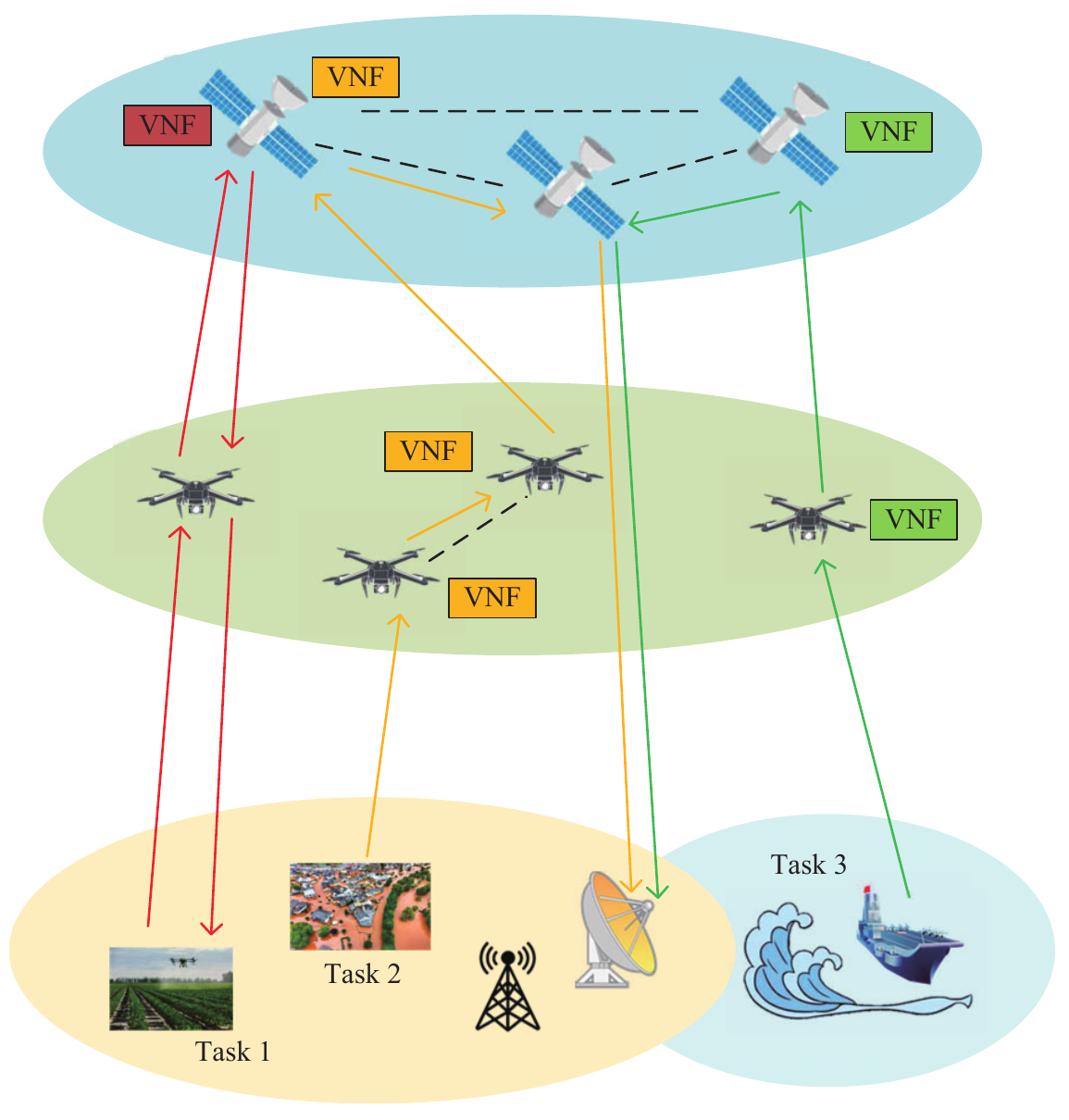}}
    \caption{A scenario of SFC deployment in SAGIN.}
    \label{fig1}
\end{figure}

As illustrated in Fig. \ref{fig1}, a SAGIN scenario is composed of users on the ground, unmanned air vehicles (UAVs) in the air and LEO satellites in the space. Terrestrial facilities include data centers, base stations, satellite receivers, etc. Users include mobile facilities on the land, ships in the ocean, etc. LEO satellites are interconnected by wireless links, and can provide large coverage. UAVs are interconnected and can be connected to the ground base station via wireless links. Tasks come from the ground or the ocean. Then, they are served by base stations, satellite stations, UAVs, and LEO satellites. The SFC corresponding to each task is orchestrated by VNFs according to requirements. Fig. \ref{fig1} shows the orchestration and composition of SFCs for three different types of tasks, and the corresponding resource providers. However, how to depict the dynamic resources in SAGIN and the multiple tasks are still intractable.\vspace{-2mm}

\subsection{RTEG Model}

\begin{figure}[!t]
    \centerline{\includegraphics[height=10.3cm,width=8.5cm ]{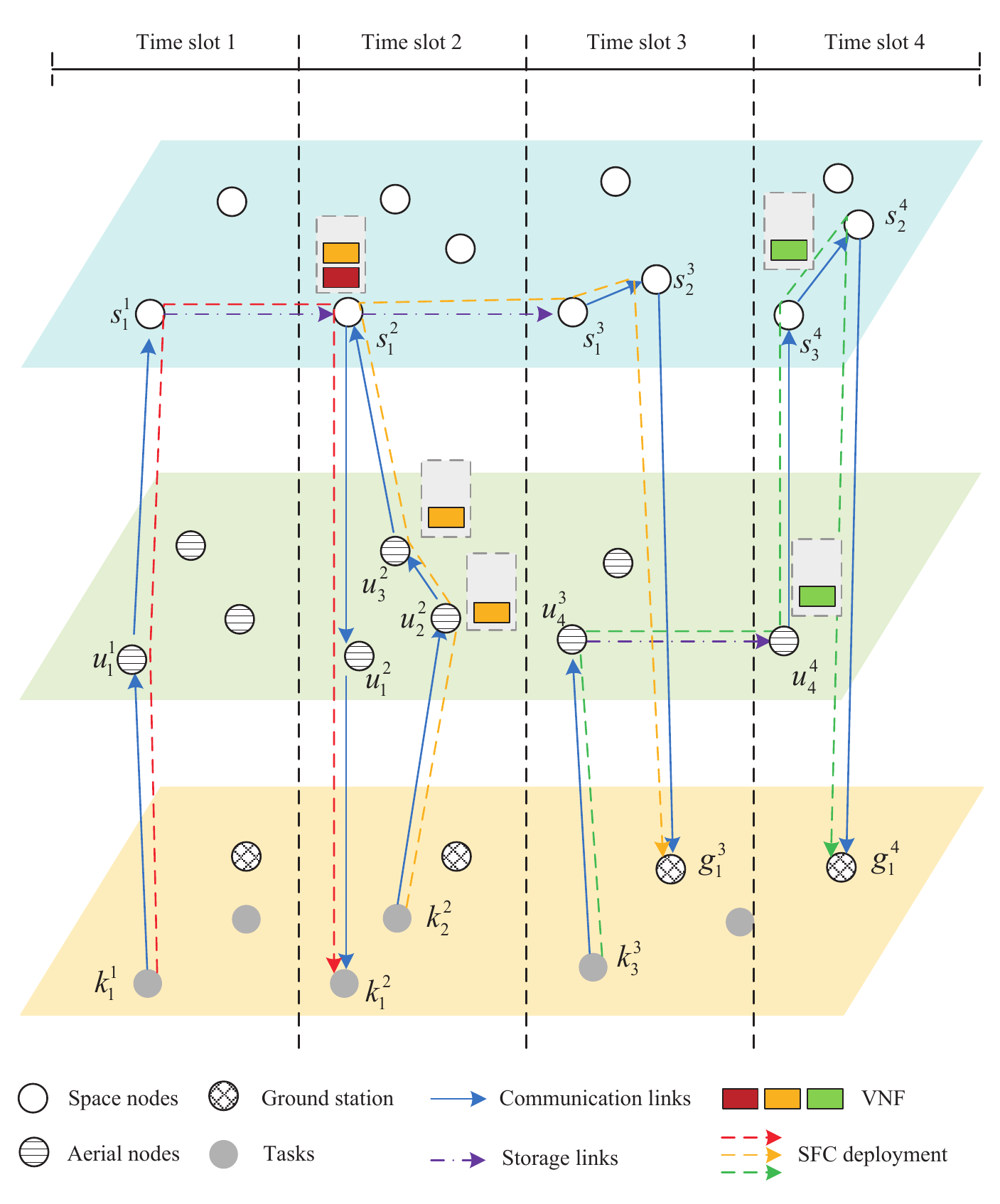}}
    \caption{A scenario of SFC deployment in SAGIN based on RTEG.}
    \label{fig2}
\end{figure}

In SAGIN, there exists high-speed periodic movement of LEO satellites, flexible motion of UAVs and low-speed movement of the ground tasks. The types of tasks are diverse, and the network is cross-domain and highly dynamic. The communication links of ground-to-UAV (G2U), UAV-to-UAV (U2U), UAV-to-satellite (U2S), satellite-to-satellite (S2S), and satellite-to-ground (S2G) change with time. The efficient deployment of SFC in such a complex cross-domain network requires effective resource characterization. Therefore, we propose the RTEG to denote the multi-layer dynamic resources of SAGIN, as shown in Fig. \ref{fig2}. 

To be specific, RTEG is characterized as $\mathcal{G} =(\mathcal{N},\mathcal{L} )$, including nodes $\mathcal{N} =\mathcal{N} _g \cup \mathcal{N} _s \cup \mathcal{N} _u$, and links $\mathcal{L} =\mathcal{L}_{gu} \cup \mathcal{L}_{uu} \cup \mathcal{L}_{us} \cup \mathcal{L}_{ss} \cup \mathcal{L}_{sg} \cup \mathcal{L}_{t}$, indicating G2U, U2U, U2S, S2S, and S2G, respectively. $\mathcal{L}_{t}=\{(n^\tau,n^{\tau+1}) | n^\tau \in \mathcal{N} _u \cup \mathcal{N} _s,1\leq \tau \leq T\}$ denotes the storage link of the same node $n$ from time slot $\tau$ to the adjacent time slot $\tau$$+1$. We consider a time horizon $\mathcal{T}$ which is larger than the period of LEO satellites. $\mathcal{T}$ is divided into $T$ time slots and the length of each time slot is $t$, $\tau \in T$. Due to the short time slot $\tau$, we consider the connections between two nodes in a same time slot are unchanged and the network topology is quasi-static. $\mathcal{N}$ represents all types of nodes, $n^\tau \in \mathcal{N}$. $\mathcal{L}$ depicts all types of links, $(n^\tau,m^{\tau'}) \in \mathcal{L}$. 

\textcolor{black}{Based on RTEG, SFC with different requirements from ground or ocean can be deployed. Corresponding to the SFC deployment scenario of SAGIN in Fig. \ref{fig1}, the deployment of three SFCs from three different tasks is depicted in Fig. \ref{fig2}. In detail, task $k^1_1$ is transmitted to $u^1_1$ and arrives at $k^2_1$ through $s^1_1$, $s^2_1$ and $u^2_1$. The VNF deploys on $k^2_1$. Task $k^2_2$ deploys the first VNF at $u^2_2$, the second VNF at $u^2_3$, and the third VNF at $s^2_1$. Then, $k^2_2$ is transmitted to the ground $g^3_1$ relayed by $s^3_1$ and $s^3_2$. Task $k^3_3$ is transmitted to $u^3_4$. It deploys the first VNF at $u^4_4$, and the second VNF at $s^4_3$. Then, it is transmitted to $g^4_1$ passing through $s^4_2$. Note that two different VNFs are deployed in $s^2_1$ and share the node resources. Therefore, we need to optimize the deployment of SFC to avoid inefficient resource utilization and obtain the  optimal deployment scheme.}

\subsection{Channel Model}

In Fig. \ref{fig1}, there exist five types of channels, including
G2U, U2U, U2S, S2S, and S2G. Since the channel types of U2U, U2S, and S2S are all related to line-of-sight communication, they are unified as air-to-air (A2A). Hence, the channels are characterized by three different models.

\subsubsection{\textcolor{black}{Channel Model of G2U}}

Following \cite{9453860}, since the antenna height of the UAV is much higher than the ground, the signal to noise ratio (SNR) of G2U can be expressed as 
\begin{equation}
    {SNR}_{gu}=\frac{P_{gu}}{N_0}(\frac{\eta _L-\eta _{NL}}{1+\alpha {\rm exp}\{-\beta (\sigma_{nm}^\tau-\alpha  )\}}+\mathcal{A}_{nm}^\tau),
\end{equation}
where
\begin{equation}
    \mathcal{A}_{nm}^\tau=20{\rm log_{20}}(\frac{4\pi d_{nm}^\tau f_c}{300})+\eta _{NL},
\end{equation}
and
\begin{equation}
    \sigma_{nm}^\tau=\frac{180}{\pi}{\rm arcsin}(\frac{h_{m^\tau}}{d_{nm}^\tau}),
\end{equation}
with $P_{gu}$ denoting the transmission power from the ground station to the UAV. $N_0$ is the power of additive white Gaussian noise. $\eta _L, \eta _{NL}, \alpha,$ and $\beta $ are constant parameters related to environments. $d_{nm}^\tau$ represents the distance between the ground station transmitter $n$ and the UAV receiver $m$ in time slot $\tau$. $h_{m^\tau}$ means the height of UAV $m^\tau$. $f_c$ denotes the carrier frequency (in MHz). 

\subsubsection{\textcolor{black}{Channel Model of A2A}}

Since the connections between UAVs and satellites are line-of-sight, the A2A channel is almost an ideal model, and related to the distance between two nodes\cite{9184929}. The SNR of A2A is \vspace{-1mm}
\begin{equation}
{SNR}_{aa}=\frac{P_{aa}G_{aa}^{tr}G_{aa}^{re}c}{4\pi f d_{aa}^\tau k_{B}T_{s}B_{aa}},
\end{equation}
where $P_{aa}$ means the transmission power. $G_{aa}^{tr}$ and $G_{aa}^{re}$ represent the transmitting and receiving antenna gains, respectively. $c$ is the speed of light. $f$ represents the carrier frequency and $d_{aa}^\tau$ indicates the distance between the sender and receiver in time slot $\tau$. $k_B$ is the Boltzmann's constant (in J/K) and $T_s$
indicates the noise temperature of the total system (in K). $B_{aa}$ is the bandwidth. 

\subsubsection{Channel Model of S2G}

The S2G channel is influenced by atmospheric precipitation, and the
S2G channel state can be predicted by the meteorological satellites\cite{9583591}.
In detail, the SNR of S2G is expressed as: \vspace{-1mm}

\begin{equation}
SNR_{sg}=\frac{P_{sg}G_{sg}^{tr}G_{sg}^{re}L_{s}L_{r}}{N_{0}B_{sg}},
\end{equation}
where $P_{sg}$ is the transmission power by S2G. $G_{sg}^{tr}$ represents the LEO satellite's transmitter antenna gain. $G_{sg}^{re}$ means the ground station's receiver antenna gain. $L_s$ is the free space loss. Since the distance between LEO satellites and ground stations is large and can be considered as unchanged in one time slot, $L_s$ is a fixed value. $L_r$ means the rain attenuation. $B_{sg}$ is the bandwidth of S2G. 

According to Shannon formula, the maximum data rate of G2U, A2A and S2U can be combined into  
\begin{equation}
r_{(n^\tau,m^\tau) }=\!B\mathrm{log}{}_{2}(1+{SNR}),\forall (n^\tau,m^\tau) \in\mathcal{L}_{gu} \cup \mathcal{L}_{aa} \cup \mathcal{L}_{sg},
\end{equation}
where $r_{(n^\tau,m^\tau)}$ denotes $r^{gu}_{(n^\tau,m^\tau)}, r^{aa}_{(n^\tau,m^\tau)},$ and $r^{sg}_{(n^\tau,m^\tau)}$. $B=B_{gu} \cup B_{aa} \cup B_{sg}$ indicates the bandwidth of G2U, A2A and S2G, respectively. $SNR$ represents ${SNR}_{gu}$, ${SNR}_{aa}$ and ${SNR}_{sg}$. $\mathcal{L}_{aa}=\mathcal{L}_{uu} \cup \mathcal{L}_{us} \cup \mathcal{L}_{ss}$.



\subsection{\textcolor{black}{\normalsize{}Energy Cost Model}}

\subsubsection{Energy Cost of UAVs}

The energy consumption of UAVs mainly includes hovering, moving, and communication\cite{9865119}. The moving power is 
\begin{equation}
    P^{\rm MOV}_{n^\tau}=\frac{v_{n^\tau}}{v^{\rm max}_{n^\tau}}(P^{\rm MAX}_{n^\tau}-P^{\rm HOV}_{n^\tau}),
\end{equation}
$\forall n^\tau\in\mathcal{N}_{u},\tau\in T$, where $v_{n^\tau}$ and $v^{\rm max} _{n^\tau}$ represents the moving speed of the UAV $n^\tau$ and the maximum speed, respectively. $P^{\rm MAX}_{n^\tau}$ and $P^{\rm HOV}_{n^\tau}$ denotes the power at the UAV's maximum speed and the hovering power, respectively. The hovering power can be written as
\begin{equation}
    P ^{\rm HOV}_{n^\tau}=\Delta  \sqrt{\frac{(M_{n^\tau}) ^3}{\mu ^2_{n^\tau} \nu _{n^\tau}}} ,\forall n^\tau\in\mathcal{N}_{u},\tau\in T,
\end{equation}
where $\Delta =\sqrt{{g^3}/{(2\pi\vartheta) }}$ denotes the environmental parameter. $g$ is the Earth's gravity acceleration. $\vartheta $ means the air density. $M_{n^\tau}$ represents the mass. $\mu_{n^\tau} $ and $\nu_{n^\tau} $ indicates the radius and the number of propellers in the UAV $n^\tau$, respectively. Hence, the path energy cost is
\begin{equation}
    E_{n^\tau,u}^{\rm PATH}=P^{\rm MOV}_{n^\tau} \frac{\Vert \gamma _{n^\tau}-\gamma _{n^{\tau+1}}\Vert _2}{v_{n^\tau}}+P ^{\rm HOV}_{n^\tau}t,
\end{equation}
$\forall n^\tau\in\mathcal{N}_{u},\tau\in T$, where $\gamma _{n^\tau}$ indicates the position of the UAV $n$ in the time slot $\tau$. Besides, the communication energy cost can be expressed as\vspace{-2mm}
\begin{equation}
    E_{n^\tau,u}^{\rm COM}=\underset{k\in\mathcal{K}}{\sum}\underset{m^\tau \in\mathcal{N}}{\sum}\frac{P^{tr}_{n^\tau}z^k_{(n^\tau,m^\tau)} \varphi _k}{r_{(n^\tau,m^\tau) }},
\end{equation}
$\forall n^\tau\in\mathcal{N}_{u},\tau\in T$, where $P^{tr}_{n^\tau}$ is the transmitted power of the UAV $n^\tau$. $z^s_{(n^\tau,m^\tau)}\in \{0,1\} $ indicates whether the SFC of task $s$ is deployed on the link $(n^\tau,m^\tau)$. $\varphi _s$ represents the amount of communication resource required by task $s$ in each SFC. Therefore, the total energy cost is \vspace{-3mm}

\begin{equation}
E_{n^\tau,u}^{total}=E_{n^\tau,u}^{\rm PATH}+E_{n^\tau,u}^{\rm COM},\forall n^\tau\in\mathcal{N}_{u},\tau\in T, \label{formula:energycost of HAP}
\end{equation}

\subsubsection{\textcolor{black}{Energy Cost of LEO satellites}}

The energy cost of LEO satellites is mainly related to the transmission and reception of data. The energy cost is denoted as $E_{n^\tau,s}^{re}$ when the satellite is the receiver (U2S, S2S), and the energy cost is depicted as $E_{n^\tau,s}^{tr}$ when the satellite is the transmitter (S2S, S2G). We have
\begin{equation}
    E_{n^\tau,s}^{re}\!=\!\underset{k\in\mathcal{K}}{\sum}\!\left(\underset{m^\tau \in\mathcal{N}_{u}}{\sum}\!\!\frac{P_{us}^{re}z^k_{(m^\tau,n^\tau)} \varphi _k}{r^{aa}_{(m^\tau,n^\tau) }}\!+\!\!\!\underset{m^\tau\in\mathcal{N}_{s}}{\sum}\!\!\frac{P_{ss}^{re}z^k_{(m^\tau,n^\tau)} \varphi _k}{r^{aa}_{(m^\tau,n^\tau) }}\!\right)\!,\label{formula: Energycost LEO}
\end{equation}
and
\begin{equation}
    E_{n^\tau,s}^{tr}\!=\!\underset{k\in\mathcal{K}}{\sum}\!\left(\underset{m^\tau \in\mathcal{N}_{s}}{\sum}\!\!\frac{P_{ss}^{tr}z^k_{(n^\tau,m^\tau)} \varphi _k}{r^{aa}_{(n^\tau,m^\tau) }}\!+\!\!\!\underset{m^\tau\in\mathcal{N}_{g}}{\sum}\!\!\frac{P_{sg}^{tr}z^k_{(n^\tau,m^\tau)} \varphi _k}{r^{sg}_{(n^\tau,m^\tau) }}\!\right)\!,
\end{equation}
$\forall n^\tau\in\mathcal{N}_{s},\tau\in T$, where $P_{us}^{re}$ and $P_{ss}^{re}$ are the receive power of U2S and S2S, respectively. $P_{ss}^{tr}$ and $P_{sg}^{tr}$ are the transmitted power of S2S and S2G, respectively. The total energy cost can be expressed as 
\begin{equation}
E_{n^\tau,s}^{total}=E_{n^\tau,s}^{re}+E_{n^\tau,s}^{tr}+E_{n^\tau,s}^{o}, \forall n^\tau\in\mathcal{N}_{s},\tau\in T,
\end{equation}
where $E_{n^\tau,s}^{o}$ is the general operation energy consumption.

\section{Problem Formulation\label{sec:Problem-Formulation}}

\textcolor{black}{In this Section, the optimization problem is formulated, including three types of constraints.}\vspace{-2mm}
\subsection{Constraints}
\subsubsection{Deployment Constraints}

Each VNF $f_r\in \mathcal{F} _k$ of task $k$ can only be deployed on one node, i.e., 
\begin{equation}
    \underset{n^\tau}{\sum} x^{k,f_r}_{n^\tau}=1,\forall k,f_r,\label{cons:x01}\vspace{-1mm}
\end{equation}
where $x^{k,f_r}_{n^\tau}\in \{ 0,1\} $ indicates whether VNF $f_r$ of task $k$ is deployed to node $n^\tau$ in time slot $\tau$. Also, VNF $f_r$ can be deployed to node $n^\tau $ only if the SFC of task $k$ passes through the node, i.e., 
\begin{equation}
    x^{k,f_r}_{n^\tau}\leq y^k_{n^\tau},\forall k,f_r,n^\tau,\label{cons:y01}
\end{equation}
where $y^k_{n^\tau} \in \{ 0,1\} $ indicates whether SFC of task $k$ passes through node $n^\tau$ of RTEG.

The SFC for each task is one-way. When the SFC passes through node $n^\tau _i$, it is possible to deploy the SFC on a link where the node is connected to surrounding nodes, including the connection of the same node between adjacent time slots. Hence, the constraint can be expressed as
\begin{equation}
    \underset{(n^\tau,m^\tau)}{\sum}z^k_{(n^\tau,m^\tau)}+z^k_{(n^\tau,n ^{\tau+1})}\leq y^k_{n^\tau},\forall k, n^\tau,m^\tau,\label{cons:z01}
\end{equation}

\subsubsection{Resource Constraints}

The storage resource of a node can be transformed into the link resource of the same node between adjacent time slots in RTEG. The storage resource capacity of the node $G_{n^\tau} (n^\tau\in\mathcal{N}) $, is limited, characterized as the link capacity of the node across time slots in RTEG. Thus, the total storage resource on the node $n^\tau$ cannot exceed $G_{n^\tau}$, i.e.,
\begin{equation}
    \sum_{k\in \mathcal{K}}^{}  z^k_{(n^\tau,n ^{\tau+1} )} \delta^k_{n^\tau}  \leq G_{n^\tau},\forall (n^\tau,n ^{\tau+1}), \tau \in T,\label{cons:storageCapacity}
\end{equation}
where $\delta^k_{n^\tau}$ means the amount of storage resource required by task $k$ on the node $n^\tau$.

Since the computing resource capacity of each UAV and LEO satellite is limited, the total computing resource consumed by SFC deployment on the node cannot exceed $C^u_ {n^\tau} (n^\tau\in\mathcal{N}_u)$ and $C^s_{n^\tau} (n^\tau\in\mathcal{N}_s)$, i.e.,
\begin{equation}
    \sum_{k\in \mathcal{K} }^{}  \sum_{f\in \mathcal{F} _r}^{} x^{k,f_r}_{n^\tau} \sigma ^{f_r}_k\leq C_ {n^\tau} ,\forall n^\tau\in\mathcal{N}_u \cup \mathcal{N}_s, \tau \in T,\label{cons:comcapacity1}
\end{equation}
where $\sigma ^{f_r}_s$ means the amount of computing resource required by VNF $f_r$ of task $k$. $C_ {n^\tau}$ represents $C^u_ {n^\tau} (n^\tau\in\mathcal{N}_u)$ and $C^s_{n^\tau} (n^\tau\in\mathcal{N}_s)$. Besides, the computing resource consumption $E^C_{n^\tau}$ can be expressed as
\begin{equation}
    E^{C}_{n^\tau,u}=\sum_{k\in \mathcal{K} }^{}  \sum_{f\in \mathcal{F} _k}^{}  x^{k,f_r}_{n^\tau} \sigma ^{f_r}_k e^c, \forall n^\tau\in\mathcal{N}_u \cup \mathcal{N}_s, \tau \in T,\label{cons:energycom1}
\end{equation}
where $E^{C}_{n^\tau}$ includes $E^{C}_{n^\tau,u}$ and $E^{C}_{n^\tau,s}$. $e^c$ is consisted of $e^c_u$ and $e^c_s$, which are the energy cost per unit of computing resource on a UAV and a LEO satellite, respectively.
The energy consumption of a node includes computing resource consumption $E^C_{n^\tau}$ and transmission resource consumption $E^{total}_{n^\tau }$. The total energy consumption can not exceed the total energy of the node $E^{MAX}$, i.e.,
\begin{equation}
    E^C_{n^\tau}+E^{total}_{n^\tau}\leq E^{MAX}_{n^\tau} ,\forall {n^\tau} \in \mathcal{N}_u \cup \mathcal{N}_s,\tau \in T,\label{cons:energytotal}
\end{equation}

As mentioned above, different channels have different communication capabilities. In one time slot, the data amount for all tasks on link $(n^\tau,m^\tau)$ cannot exceed its channel capacity, i.e.,\vspace{-1mm}
\begin{equation}
    \sum_{k\in \mathcal{K}}^{}  z^k_{(n^\tau,m^\tau)} \varphi _k\leq r_{(n^\tau,m^\tau)}t,\forall (n^\tau,m^\tau)\in \mathcal{L}, \tau\in T.\label{cons:links1}
\end{equation}

\subsubsection{Flow Constrains}

There exist some flow conservation constraints which should be satisfied, i.e.,
\begin{align}
    &\sum_{(o^\tau,m^\tau)} z^k_{(o^\tau,m^\tau)}=1,\forall k,\tau,\label{cons:flowconsOD1}\\
    &\sum_{(n^\tau,d^\tau)} z^k_{(n^\tau,d^\tau)}=1,\forall k,\tau,\label{cons:flowconsOD2}\\
    &\sum_{(n^\tau,m^\tau)}^{} \!z^k_{(n^\tau,m^\tau)}+\!\!\sum_{(m^{\tau-1},m^\tau)}^{} \!z^k_{(m^{\tau-1},m^\tau)}\!=\! \nonumber\\
    &\sum_{(m^\tau,n^\tau)}^{} \!z^k_{(m^\tau,n ^\tau)}+\!\!\sum_{(m^\tau,n^{\tau+1})}^{} \!z^k_{(m^\tau,n^{\tau+1})},\forall k,\tau\neq 1 or T,m^\tau,  \label{cons:flowconserS}
\end{align} 
where $o^\tau$ means the original node and $d^\tau$ denotes the destination node. $z^k_{(m^{\tau-1},m^\tau)}=0$ if $\tau=1$, and $z^k_{(m^\tau,n^{\tau+1})}=0$ if $\tau=T$.\vspace{-2mm}

\subsection{Optimization Objective}

The objective problem is to maximize the number of tasks which complete SFC deployment, indicated as $\mathbf{Q}$ , i.e.,
\begin{align}
\mathscr{P}0:\;\underset{\boldsymbol{X},\boldsymbol{Y},\boldsymbol{Z}}{\textrm{max}}\; \mathbf{Q} \\
 \textrm{s.t.}\quad & (\ref{cons:x01})-(\ref{cons:flowconserS}),
\end{align}
where $\boldsymbol{X}=\{x^{k,f_r}_{n^\tau},\forall n^\tau \in \mathcal{N},\tau\in T\} $, $\boldsymbol{Y}=\{y^k_{n^\tau},\forall n^\tau \in \mathcal{N},\tau\in T\}$, and $\boldsymbol{Z}=\{z^k_{(n^\tau,m^\tau)},\forall (n^\tau,m^\tau)\in\mathcal{L},\tau\in T\}$. 
Note that $\mathscr{P}0$ is a MILP problem, which is hard to solve in limited time complexity. Therefore, we use the matching game to deal with it.

\vspace{-2mm}
\section{Algorithm Design\label{sec:Algorithm-Design}}

\begin{algorithm}[t]
    \caption{MG-RTEG.\label{Matching algorithm}}

    \begin{algorithmic}[1]
        \REQUIRE All tasks $k \in \mathcal{K} $ with the set of SFCs, all nodes $n \in \mathcal{N} $, the data size of tasks $d_k$.
        \ENSURE Number of tasks completing the SFC deployment.
        \STATE Use the Dijkstra's algorithm to obtain the shortest paths of tasks from the starting point to the destination.
        \STATE Put nodes of the shortest path into task's preference list $L_k$ orderly.
        \STATE Sort tasks which arrive at $n$ according to $d_k$ in a descending order.
        \STATE Put tasks into the node's preference list $L_n$ in sequence.
        \FOR{each time slot $\tau$} 
        \FOR{there exists $k$ not transmitted}
        \IF{$d_k$ is small enough to be transmitted entirely by UAVs}
            \FOR {$n^\tau \in L_k$}
                \STATE Allocate VNFs of $k \in L_n$ to $n^\tau$ until no more VNFs can be deployed.
                \STATE Reject all $k'$ that satisfied $k \succ_n k'$.
                \STATE Remove $n^\tau$ from $L_k$.           
            \ENDFOR
            \IF{$k$ fails to deploy any VNF in $\tau$}
            \STATE Find $n_0 \in L_k$ closest the starting point.
            \STATE Store $k$ to the next time slot at $n^{\tau+1}_0$.
            \ENDIF
            \IF{$k$ reaches the node closest to the destination and VNFs are not entirely deployed}
            \STATE Find $n_1 \in L_k$ which is the last node that deploys VNFs.
            \STATE Store $k$ to the next time slot at $n^{\tau+1}_1$.
            \ENDIF
        \ELSE
        \STATE $k$ transmits to satellites via a UAV nearest the starting point and deploys VNFs on satellites.
        \ENDIF
        \ENDFOR
        \ENDFOR
    \end{algorithmic}
\end{algorithm}

We deal with this MILP problem by Algorithm \ref{Matching algorithm}, which named matching game-RTEG (MG-RTEG). This problem is indicated as a many-to-one two-sided matching game problem\cite{gale1962college,9174937,9714482}. It can be solved by the Gale-Shapley (GS) algorithm, also known as the deferred-acceptance algorithm. The GS algorithm has an advantage over the matching proposer, so we set tasks as the matching proposers, while the UAV and satellite nodes as the matching receivers. 

Firstly, we build the preference lists, which represent nodes that the task is more preferred to select and tasks that the node likes better. The preference list of the task is constructed by the Dijkstra's algorithm. It is used to generate the shortest path of the task from the starting point to the destination. The nodes in the obtained path are put into the preference list in sequence from the UAV nodes close to the starting point (line 1-2). The preference list of the node is constructed according to the data size of tasks in a descending order (line 3-4). Then, we obtain the deployment of SFC by the GS algorithm. In the selection of nodes for task transmission, we consider priority to UAV nodes, which are close to the starting point of tasks, and have high flexibility. If the data size of the task is too large to be transmitted totally by the UAVs, or the computing resources of the UAV nodes are exhausted, the deployment of SFC is considered to be transmitted to satellites (line 22).

During one time slot, the task selects the first node according to the preference list and deploys VNFs. The node selects the corresponding task according to the preference list and rejects the other tasks behind it in the preference list. According to the delayed acceptance of the GS algorithm, the node can reject the currently task $s'$ and accept the task $k$ that ranks higher in the node's preference list, indicated as $k \succ_n k'$. In particular, $k \succ_n k'$ denotes that the node $n$ prefers task $k$ than task $k'$. Furthermore, once the task is rejected, tasks that rank behind it are also rejected (line 5-12). If there is a task that fails to deploy any VNF in a time slot, the task is stored at the UAV node closest to the starting point. It waits at this node until the next time slot to complete the matching comparing with other tasks (line 13-16). If there exists a task that only deploys a part of VNFs, the task is stored at the node where VNFs are last deployed, waiting until the next time slot to deploy the remaining VNFs (line 17-20). The iteration is carried out until the two-sided matching achieves stability. Then, SFC deployment is completed.\vspace{-1mm}

\section{Simulation Results\label{sec:Simulation-Results}}

\begin{figure}[!t]
    \centering
    
    \subfloat[The number of tasks which complete SFC deployment.\label{fig3}]{
    
    \includegraphics[height=5.8cm,width=7.35cm ]{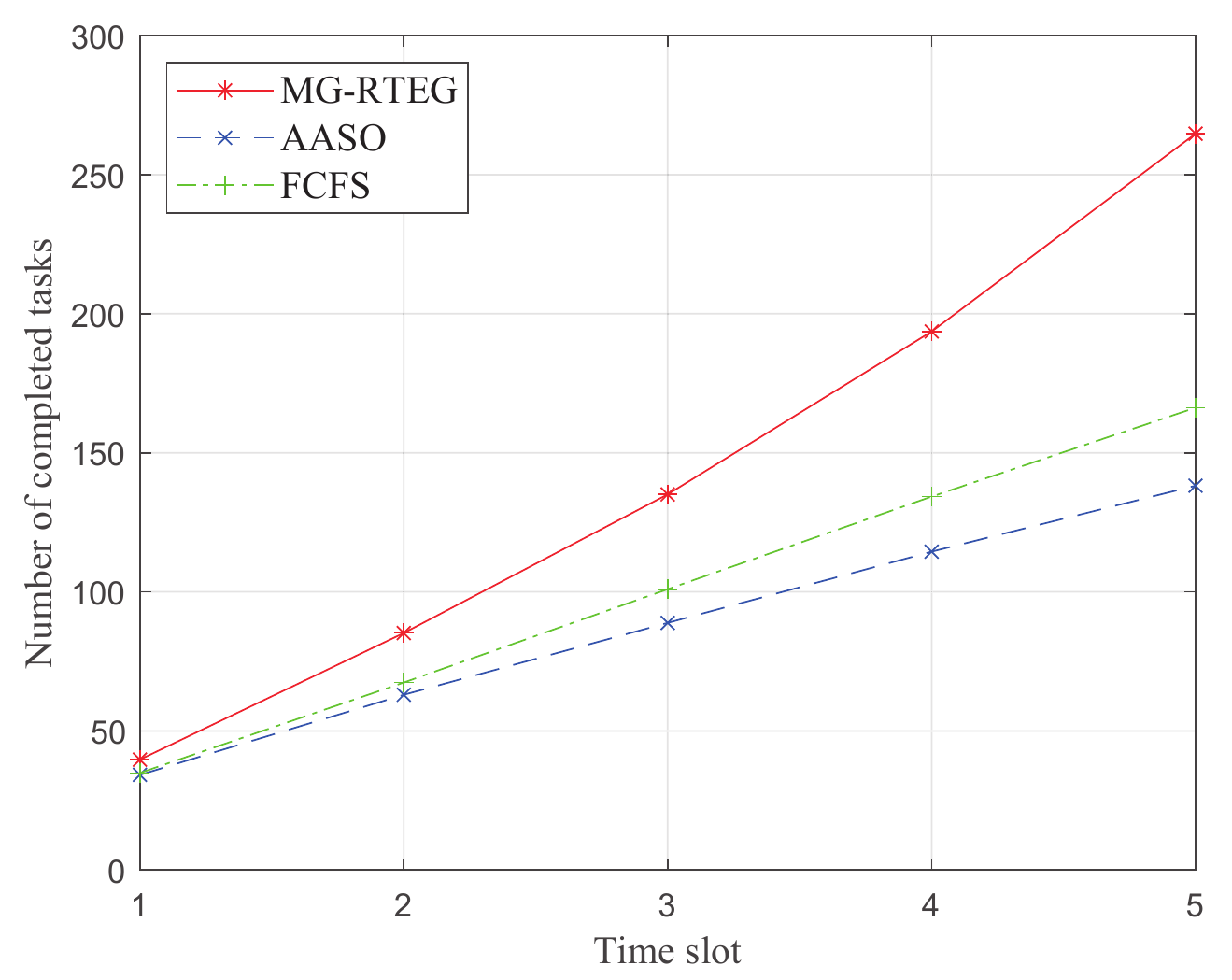}
    }
    
    \subfloat[Nodes' computing resource utilization.\label{fig4}]{
    
    \includegraphics[height=5.8cm,width=7.25cm]{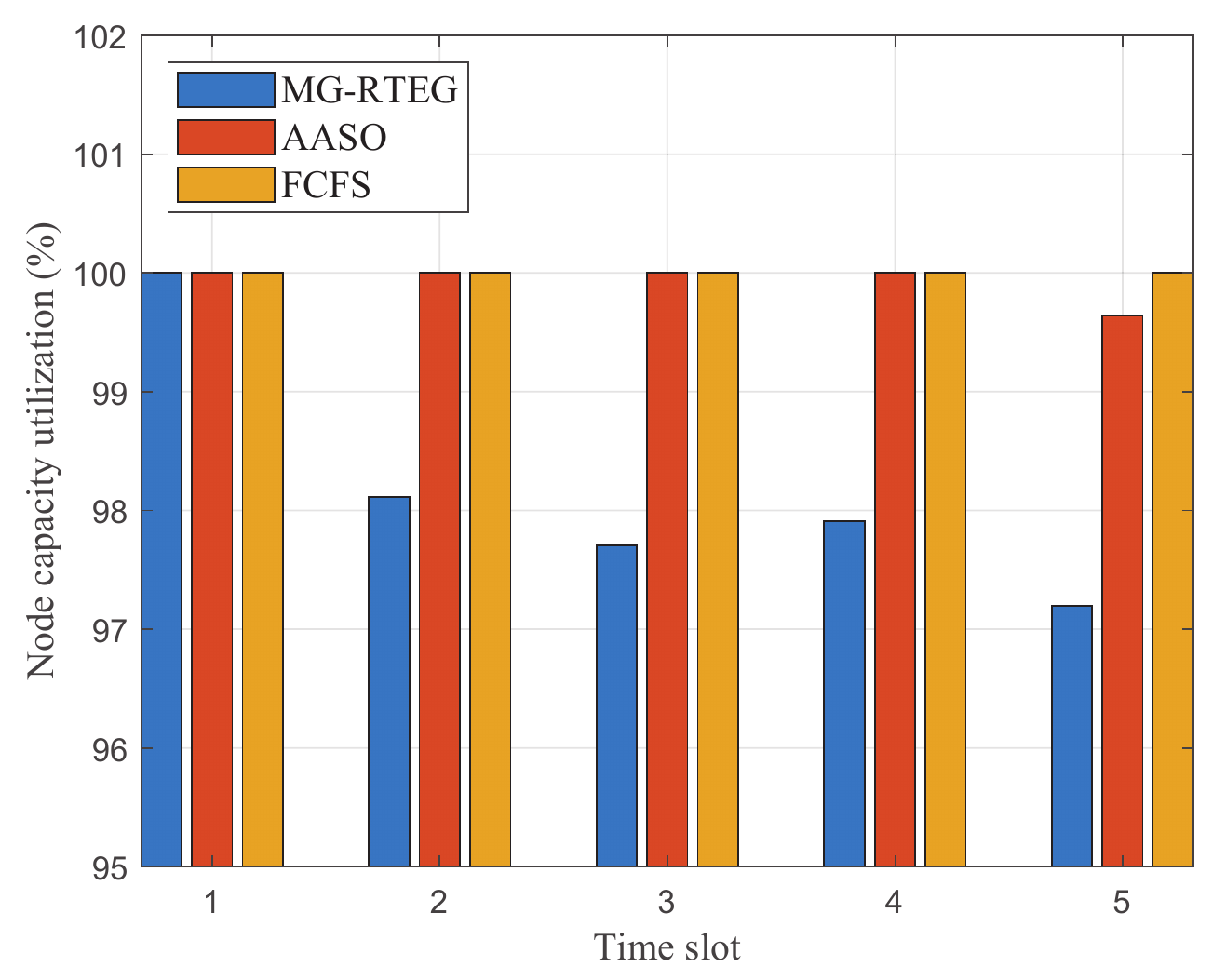}
    }
    
    \caption{Comparison of different algorithms in a time slot.\label{fig:Comparison}}
    
    \vspace{-5mm}
\end{figure}

We carry out simulation designs by MATLAB and set data as six UAVs and one satellite. We assume that these UAVs can always communicate with the nearest LEO satellite, ignoring the switching issue. UAVs and satellites are dynamically changing in each time slot. The computing resources of each UAV and satellite are set as 8 units and 50 units, respectively. We consider one VNF type of the task, $|\mathcal{F} _s| =1$. $\sigma ^{f_k}_s = 1$unit$/$10Mbit. Tasks are randomly distributed on the ground within the coverage of the UAVs. The data size of each task is randomly generated between 10Mbit to 50Mbit.

\begin{figure}[!t]
    \centering
    \centerline{\includegraphics[height=5.8cm,width=7.35cm ]{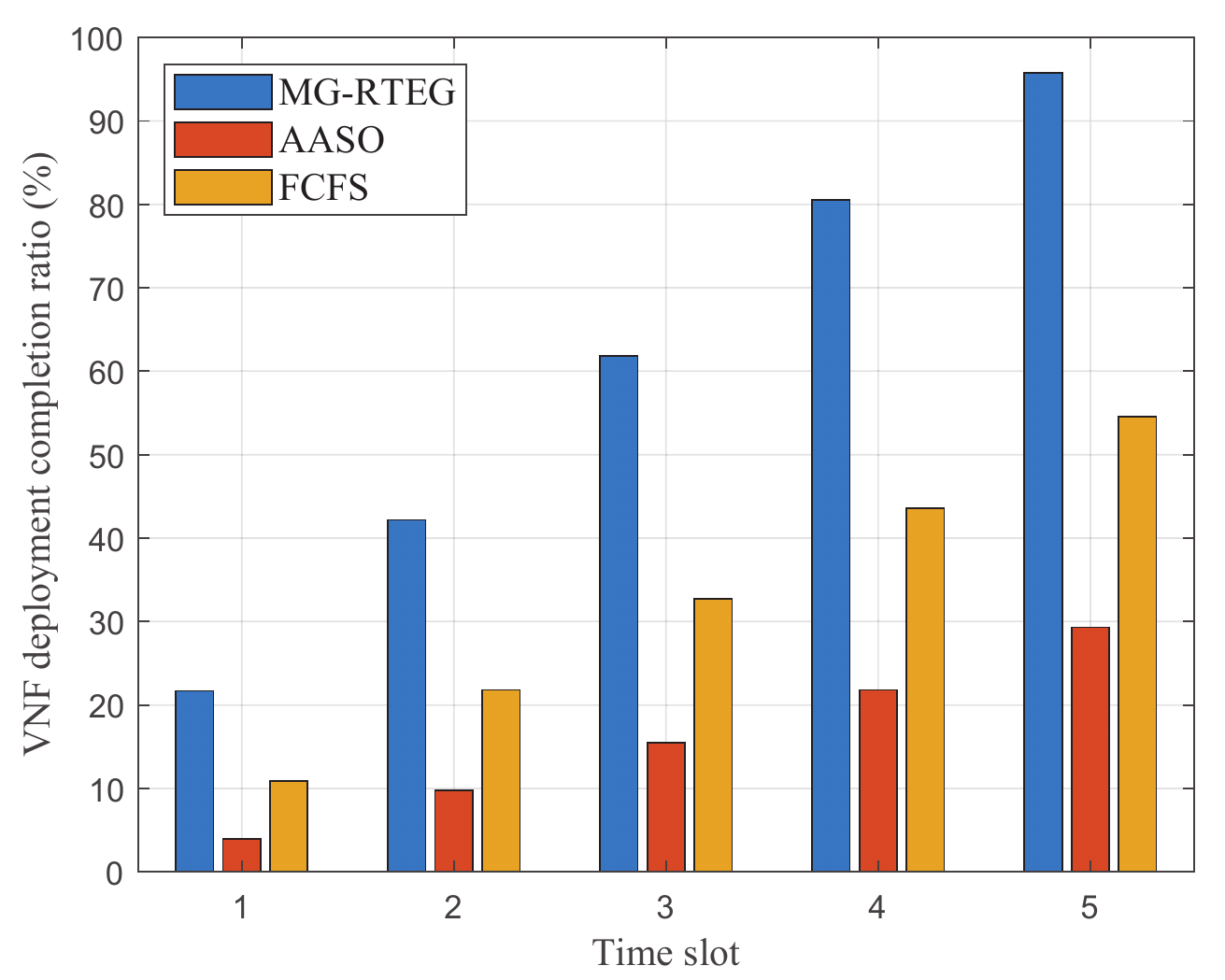}}
    \caption{Comparison of VNFs' computing resource usage required for tasks in each time slot.}
    \label{fig5}
\end{figure}

In order to evaluate the advantages of the proposed matching algorithm, we consider 300 random tasks, and compare our proposed algorithm with two other algorithms. One is selecting node's accepting tasks in an ascending order (AASO), and another is the first-come-first-service (FCFS). It is observed from Fig. \ref{fig3} that with the increment of the time slot, the number of tasks which complete SFC deployment gradually increases. As for the completed task number, the proposed algorithm is always greater than the other two algorithms in each time slot. In addition, as shown in Fig. \ref{fig4}, due to the large number of tasks, the computing resources of the nodes are almost exhausted, while the nodes are not fully used in the subsequent time slots with the proposed algorithm. Therefore, compared with the other two algorithms, MG-RTEG can successfully complete more tasks and deploy more SFCs with less node computing resource  consumption. Considering the amount of resources required for VNF deployment of all tasks in Fig. \ref{fig5}, the ratio of deployment completion is much higher by MG-RTEG. In summary, the performance of proposed algorithm is superior to the other two algorithms.

\section{Conclusions\label{sec:Conclusions}}

In this paper, we leverage RTEG to characterize the situation of different resources in SAGIN, and formulate the resource management as a SFC deployment optimization problem. Then, we propose a matching game based algorithm to solve the designed optimization problem of maximizing the number of completed tasks. Simulation results show that MG-RTEG can allocate the resources more effectively and stably, and deploy more SFCs. Meanwhile, MG-RTEG can enable high resource utilization. It shows that the proposed algorithm outperforms other compared algorithms.

\textcolor{black}
{\bibliographystyle{IEEEtran}
\bibliography{ref}
}
\end{document}